\documentclass[a4paper]{article}

\usepackage{INTERSPEECH2020}

\title{Advancing Speech Synthesis using EEG}
\name{Gautam Krishna, Co Tran, Mason Carnahan, Ahmed H Tewfik}
\address{
  Brain Machine Interface Lab, The University of Texas at Austin
  }
\email{}

\begin{document}

\maketitle
\begin{abstract}
In this paper we introduce attention-regression model to demonstrate predicting acoustic features from electroencephalography (EEG) features recorded in parallel with spoken sentences. First we demonstrate predicting acoustic features directly from EEG features using our attention model and then we demonstrate predicting acoustic features from EEG features using a two-step approach where in the first step we use our attention model to predict articulatory features from EEG features and then in second step another attention-regression model is trained to transform the predicted articulatory features to acoustic features. Our proposed attention-regression model demonstrates superior performance compared to the regression model introduced by authors in \cite{krishna2020synthesis} when tested using their data set for majority of the subjects during test time. The results presented in this paper further advances the work described by authors in \cite{krishna2020synthesis}.  
\end{abstract}

\section{Introduction}

We human beings have the unique ability to communicate with each other by producing intelligible speech but people recovering from stroke, amyotrophic lateral sclerosis (ALS) etc suffer from aphasia and other speaking disabilities which makes it challenging for them to produce intelligible speech. This also limits technology accessibility to them as they can't interact with voice activated virtual personal assistants like Amazon Alexa, Apple Siri, Samsung Bixby etc. Recently researchers have started investigating the possibility of developing neural signal based speech prosthetic to address this issue. For the example the work explained in \cite{krishna20,krishna2019state,krishna2019speech} shows speech recognition using non-invasive electroencephalography (EEG) neural signals, where EEG recorded in parallel with speech, passive listening are translated to text. In another recent work described in \cite{anumanchipalli2019speech,herff2019generating,angrick2019speech} authors demonstrated synthesizing speech directly from Electrocorticography (ECoG) signals using deep learning models. 
The ECoG is a invasive procedure where a brain surgery is performed to implant the ECoG electrodes to get the electrical recordings. On the other hand EEG is a non-invasive technique where EEG sensors are placed on the scalp of the subject to obtain EEG recordings. The ECoG demonstrates better spatial resolution and signal-to-noise ratio compared to EEG but the non-invasive nature of EEG makes it more safer and easily deployable than ECoG technology. The EEG signals also offer high temporal resolution. 

In \cite{krishna2020synthesis} authors provided preliminary results for synthesizing speech directly from EEG features. They demonstrated predicting acoustic features directly from EEG features using various EEG feature sets introduced by authors in \cite{krishna2019state} using a gated recurrent unit (GRU) \cite{chung2014empirical} based regression model. In their regression model their encoder GRU hidden states outputs had one-to-one mapping with their time distributed dense layer decoder. A similar principle was also used by authors in \cite{anumanchipalli2019speech} in designing their decoder but they used a different recurrent neural network (RNN) model instead of GRU.
In this paper we make use of encoder-decoder model with attention mechanism \cite{luong2015effective} to perform regression where a luong dot product attention layer \cite{luong2015effective} present between the encoder and decoder RNN in our regression model provide context vectors to decoder RNN at every time step. The attention models have been used before for tasks like automatic speech recognition (ASR) \cite{chorowski2015attention} and speech synthesis (producing sound from text) \cite{wang2017tacotron} where the inputs and outputs are of different lengths, whereas in our case the input and output features are of same length since both EEG and speech signals are recorded simultaneously. 

In this paper we first demonstrate predicting acoustic features directly from EEG features using our attention model and then we demonstrate predicting acoustic features from EEG features using a two-step approach where in the first step we use our attention model to predict articulatory features from EEG features and then in second step another attention-regression model is trained to transform the predicted articulatory features to acoustic features. Our proposed attention-regression model demonstrates superior performance compared to the regression model introduced by authors in \cite{krishna2020synthesis} when tested using their data set for majority of the subjects during test time. The results presented in this paper further advances the work described by authors in \cite{krishna2020synthesis}.

\section{Attention-Regression Speech Synthesis model}

The architecture of the attention-regression model is described in Figure 1. Our encoder model is a GRU with 256 hidden units. The encoder GRU takes input features and transforms it into output hidden features. A luong dot product attention layer \cite{luong2015effective} takes the encoder GRU ouput hidden features and calculates the context vectors which are passed to the decoder GRU with 128 hidden units. A dropout \cite{srivastava2014dropout} regularization with dropout rate 0.2 is applied after the attention layer.  
The attention layer calculations are described below.

\begin{equation}
  c_{k} =
    \begin{cases}
      \sum_{t=1}^{T}\vec{h_t}\alpha_{k,t}
      
    \end{cases}       
\end{equation}
\begin{equation}
  \alpha_{k,t} =
    \begin{cases}
      softmax(score(\vec{h_t},\vec{h_{s-1}}))
      
    \end{cases}       
\end{equation}
\begin{equation}
  score(\vec{h_t},\vec{h_{s-1}}) =
    \begin{cases}
      W \cdot \vec{h_t}^\intercal \cdot \vec{h_{s-1}}
      
    \end{cases}       
\end{equation}
where $c_k$ is the context vector, $\alpha_{k,t} $ is the attention weight vector, $\vec{h_t}$ is hidden state output vector of the encoder GRU and $\vec{h_{s-1}}$ is hidden state of the decoder GRU at time step $k-1$. The $\alpha_{k,t}$ is a measure of how much attention $y_k$ must pay to $\vec{h_t}$, $t=\{1, 2, 3,\cdots \cdots,T\}$, $y_k$ is the prediction at time step $k$ by the decoder GRU and $T$ is the number of time steps. The weight matrix $W$ is learned during training of the model. The number of time steps of the encoder and decoder GRU are same since both the input and target features were recorded synchronously. The decoder GRU outputs are passed to a time distributed dense layer with linear activation function. The number of hidden units in the time distributed dense layer depends on the target vector dimension. For example, for the experiment involving predicting acoustic features directly from EEG features or articulatory features, the time distributed dense layer has 13 or 128 hidden units depending on number of Mel-frequency cepstral coefficients (MFCC) used and for the experiment involving predicting articulatory features from EEG features the time distributed dense layer has 6 hidden units. Overview of the experiments carried out are explained in Figure 2. 

The model was trained for 2500 epochs with adam \cite{kingma2014adam} as the optimizer. The batch size was set to 100 and the validation split hyper parameter was set to a value of 0.1. We used mean square error (MSE) as the loss function.

\begin{figure}[h]
\begin{center}
\includegraphics[height=8.5cm, width=\linewidth,trim={0.1cm 0.1cm 0.1cm 0.1cm}]{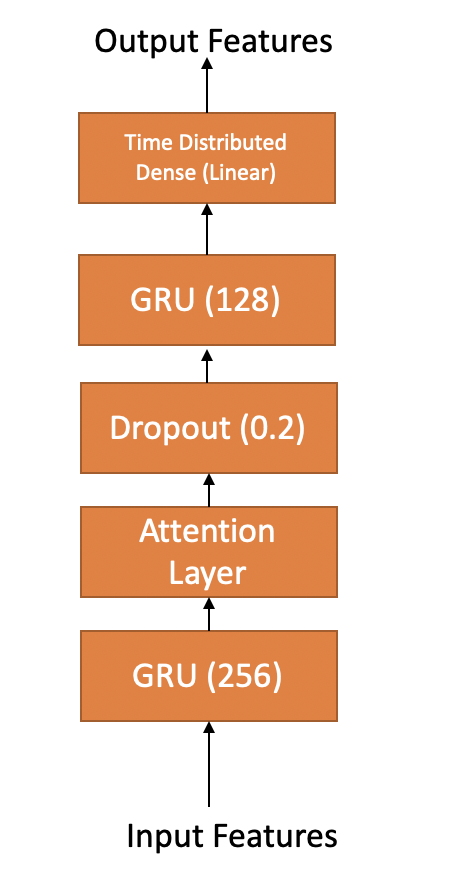}
\caption{Attention Regression Model} 
\label{1vsall}
\end{center}
\end{figure}

\begin{figure}[h]
\begin{center}
\includegraphics[height=8.5cm, width=\linewidth,trim={0.1cm 0.1cm 0.1cm 0.1cm}]{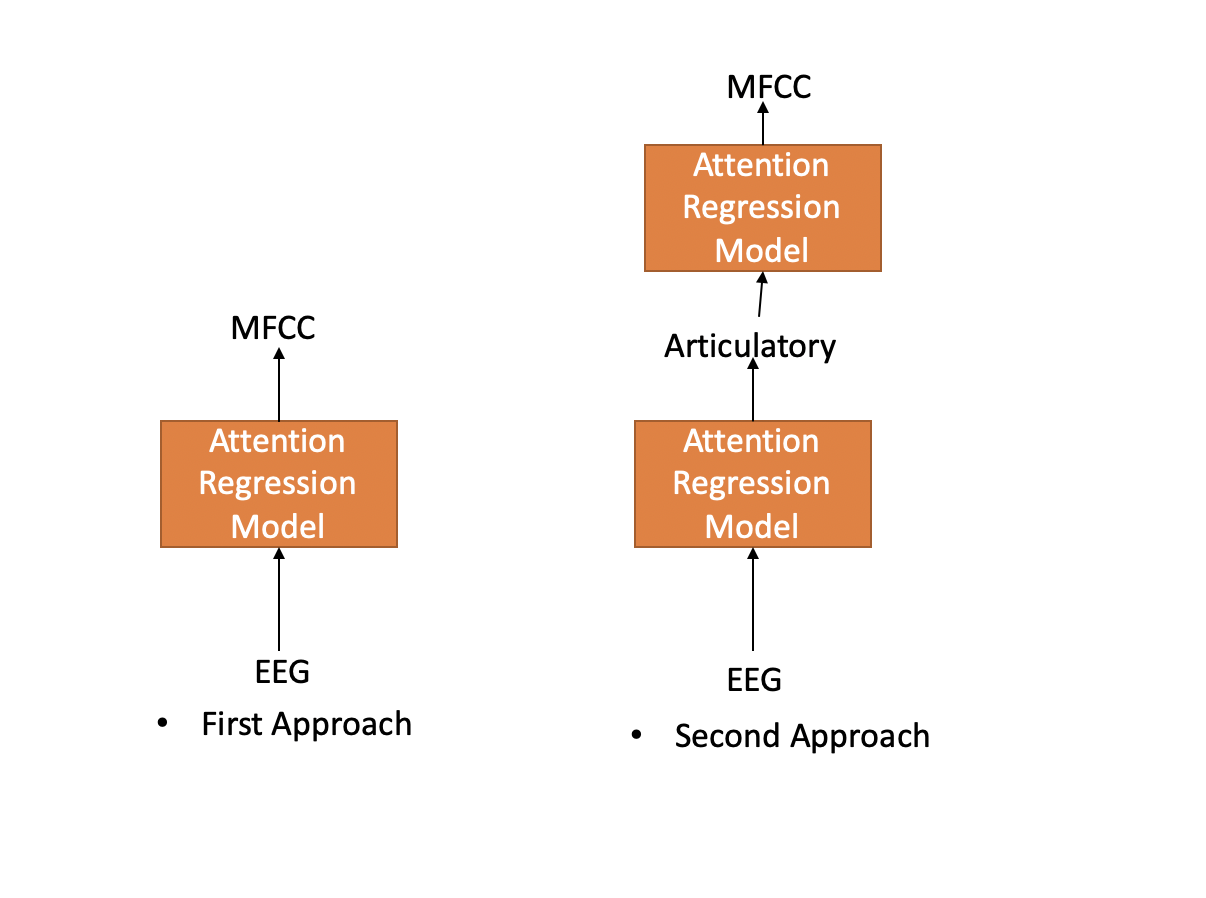}
\caption{Overview of the experiments} 
\label{1vsall}
\end{center}
\end{figure}


\section{Data Sets used for performing experiments}
We used the Data set used by authors in \cite{krishna2020synthesis} for this work. In \cite{krishna2020synthesis} authors recorded EEG signals in parallel with spoken English sentences as well as with listening utterances. In this work we use only the EEG signals recorded in parallel with spoken English sentences. 
More details of the experiment design for collecting simultaneous speech and EEG data are covered in \cite{krishna2020synthesis}. 

They used Brain product's ActiChamp EEG amplifier. Their EEG cap had 32 wet EEG electrodes including one electrode as ground. It is based on standard 10-20 EEG sensor placement method for 32 electrodes.

For each experiment set we used 80\% of the data as training set, remaining 10\% as validation set and rest 10\% as test set. The train-test split was done randomly. There was no overlap between training, testing and validation set. The way we splitted data in this work is exactly same as the method used by authors in \cite{krishna2020synthesis}. 

\section{EEG, Speech and Articulatory feature extraction details}
We followed the same EEG preprocessing methods used by authors in \cite{krishna2020synthesis}. 
The EEG signals were sampled at 1000Hz and a fourth order IIR band pass filter with cut off frequencies 0.1Hz and 70Hz was applied. A notch filter with cut off frequency 60 Hz was used to remove the power line noise.
The EEGlab's \cite{delorme2004eeglab} Independent component analysis (ICA) toolbox was used to remove other biological signal artifacts like electrocardiography (ECG), electromyography (EMG), electrooculography (EOG) etc from the EEG signals. 
We then extracted the three EEG feature sets explained by authors in \cite{krishna2019state}. The details of each EEG feature set are covered in \cite{krishna2019state}.

The recorded speech signal was sampled at 16KHz frequency. We extracted mel-frequency cepstral coefficients (MFCC) as features for speech signal. We used acoustic-to-articulatory speech inversion tool introduced by authors in \cite{seneviratne2018noise} to extract articulatory features of dimension 6 from the recorded speech signal. The six articulatory tract variables (TV's) that were extracted were Lip Aperture (LA), Lip Protrusion (LP), Tongue Body Constriction Location (TBCL), Tongue Body Constriction Degree (TBCD), Tongue Tip Constriction Location (TTCL) and Tongue Tip Constriction Degree (TTCD) \cite{seneviratne2018noise}.

In \cite{krishna2020synthesis} authors extracted MFCC 13 coefficients and extracted MFCC, EEG features for each channel at a sampling frequency of 100 Hz and even though they demonstrated lower
mel cepstral distortion (MCD) \cite{kominek2008synthesizer} during test time with 100Hz features but we observed that it is possible to get more understandable reconstructed audio if we use MFCC 128 coefficients and if all features (EEG, MFCC and articulatory) are extracted at 32 Hz sampling frequency even though the MCD values were slightly higher when we used the 32 Hz features. The lower MCD values doesn't always mean the reconstructed speech is more understandable. Hence we extracted MFCC 13, EEG all three feature sets and articulatory features at 100Hz for all the four subjects in order to compare with our baseline \cite{krishna2020synthesis} and for the subject where we observed lower MCD value with 100Hz features, we also extracted MFCC 128, EEG feature set 1 and articulatory features at 32Hz.

\section{EEG Feature Dimension Reduction Algorithm Details}

We used kernel principal component analysis (KPCA) \cite{mika1999kernel} to de-noise the EEG feature space by performing dimension reduction for each EEG feature set as explained by authors in \cite{krishna2019state,krishna2020synthesis}. 
By following the dimension reduction methods explained by authors in \cite{krishna2019state} we reduced EEG feature set 1 to a dimension of 30, EEG feature set 2 was reduced to a dimension of 50 and EEG feature set 3 was kept at original dimension of 93. More details of explained variance plots used to identify the right feature dimensions are covered in \cite{krishna2019state} under their supplementary material.

\section{Results}

We computed the mel cepstral distortion (MCD) \cite{kominek2008synthesizer} between the predicted MFCC during test time and ground truth MFCC from test set to evaluate the performance of the model on test set for each subject. The predicted and ground truth MFCC values were normalized before computing the MCD values. 
We then used the Griffin Lim reconstruction \cite{griffin1984signal} algorithm to convert the predicted test time MFCC or acoustic features to audio or speech waveform. The Tables 1,2,3 and 4 shows the test time results obtained for various subjects when we used MFCC features of dimension 13 sampled at 100Hz as targets for the regression model. As seen from the table we demonstrate results using two approaches. Like mentioned before in the first approach we directly predict the MFCC or acoustic features from the EEG features using the attention regression model whereas in the second approach we first train the regression model to predict articulatory features from EEG and then the model is trained to predict acoustic features from the predicted articulatory features. Our second approach is similar to the technique used by authors in \cite{anumanchipalli2019speech}. As seen from the Tables 1,2,3 and 4 our attention-regression method demonstrated lower MCD values or better performance during test time for subjects 1,2 and 4 compared to the method used by authors in \cite{krishna2020synthesis}. Only for subject 3, the method used by authors in \cite{krishna2020synthesis} outperformed our proposed approach. For majority of the test time experiments our method outperformed their approach. One possible explanation for why a simple regression model used by authors in \cite{krishna2020synthesis} outperformed our model for subject 3 might be that the subject 3 EEG data contains high level of noise. In presence of very high level of noise in data, it is more challenging for the attention model to learn the correct alignment \cite{kim2017joint}. 
The Figures 3 and 4 shows predicted audio waveform compared to ground truth or actual waveform for subject 1 where we used MFCC 128 features sampled at 32 Hz to reconstruct the audio waveform. The predicted waveform shown in the Figures have amplitude normalized. The Y axis in plots denotes amplitude and X axis denotes number of sample points in the audio. 
We can observe from the Figures that the predicted waveform were noisier than the ground truth waveform but applying external audio filters to the predicted waveform can help in removing the noise. As seen from Table 5 using MFCC 128 features sampled at 32 Hz resulted in higher test time MCD values compared to Table 1 but like we explained before we were able to hear more intelligible speech with MFCC 128 sampled at 32 Hz compared to MFCC 13 sampled at 100 Hz. 

Another observation we noted was results among each EEG feature sets were comparable like the ones explained in \cite{krishna2020synthesis} and the results were also comparable among the first and second approaches.

\begin{table}[!ht]
\centering
\begin{tabular}{|l|l|l|l|}
\hline
\textbf{\begin{tabular}[c]{@{}l@{}}EEG\\ Feature\\ Set\end{tabular}} & \textbf{\begin{tabular}[c]{@{}l@{}}Average\\ MCD\\ Ref {[}1{]}\end{tabular}} & \multicolumn{1}{c|}{\textbf{\begin{tabular}[c]{@{}c@{}}Average\\ MCD\\ our\\ 1st\\ Approach\end{tabular}}} & \textbf{\begin{tabular}[c]{@{}l@{}}Average\\ MCD\\ our\\ 2nd\\ Approach\end{tabular}} \\ \hline
Set 1                                                                & 0.433                                                                        & 0.443                                                                                                      & 0.45                                                                                  \\ \hline
Set 2                                                                & 0.435                                                                        & \textbf{0.325}                                                                                             & \textbf{0.329}                                                                        \\ \hline
Set 3                                                                & 0.435                                                                        & 0.45                                                                                                       & 0.46                                                                                  \\ \hline
\end{tabular}
\caption{Speech synthesis test time results for subject 1 where audio is reconstructed from predicted MFCC 13, 100Hz}
\end{table}

\begin{table}[!ht]
\centering
\begin{tabular}{|l|l|l|l|}
\hline
\textbf{\begin{tabular}[c]{@{}l@{}}EEG\\ Feature\\ Set\end{tabular}} & \textbf{\begin{tabular}[c]{@{}l@{}}Average\\ MCD\\ Ref {[}1{]}\end{tabular}} & \multicolumn{1}{c|}{\textbf{\begin{tabular}[c]{@{}c@{}}Average\\ MCD\\ our\\ 1st\\ Approach\end{tabular}}} & \textbf{\begin{tabular}[c]{@{}l@{}}Average\\ MCD\\ our\\ 2nd\\ Approach\end{tabular}} \\ \hline
Set 1                                                                & 0.856                                                                        & 0.672                                                                                                      & 0.70                                                                                  \\ \hline
Set 2                                                                & 0.847                                                                        & 0.80                                                                                                       & 0.80                                                                                  \\ \hline
Set 3                                                                & 0.841                                                                        & \textbf{0.647}                                                                                             & \textbf{0.64}                                                                         \\ \hline
\end{tabular}
\caption{Speech synthesis test time results for subject 2 where audio is reconstructed from predicted MFCC 13, 100Hz}
\end{table}

\begin{table}[!ht]
\centering
\begin{tabular}{|l|l|l|l|}
\hline
\textbf{\begin{tabular}[c]{@{}l@{}}EEG\\ Feature\\ Set\end{tabular}} & \textbf{\begin{tabular}[c]{@{}l@{}}Average\\ MCD\\ Ref {[}1{]}\end{tabular}} & \multicolumn{1}{c|}{\textbf{\begin{tabular}[c]{@{}c@{}}Average\\ MCD\\ our\\ 1st\\ Approach\end{tabular}}} & \textbf{\begin{tabular}[c]{@{}l@{}}Average\\ MCD\\ our\\ 2nd\\ Approach\end{tabular}} \\ \hline
Set 1                                                                & 0.647                                                                        & 1.07                                                                                                       & 1.071                                                                                 \\ \hline
Set 2                                                                & 0.650                                                                        & 0.934                                                                                                      & 1.09                                                                                  \\ \hline
Set 3                                                                & 0.645                                                                        & 0.96                                                                                                       & 0.967                                                                                 \\ \hline
\end{tabular}
\caption{Speech synthesis test time results for subject 3 where audio is reconstructed from predicted MFCC 13, 100Hz}
\end{table}

\begin{table}[!ht]
\centering
\begin{tabular}{|l|l|l|l|}
\hline
\textbf{\begin{tabular}[c]{@{}l@{}}EEG\\ Feature\\ Set\end{tabular}} & \textbf{\begin{tabular}[c]{@{}l@{}}Average\\ MCD\\ Ref {[}1{]}\end{tabular}} & \multicolumn{1}{c|}{\textbf{\begin{tabular}[c]{@{}c@{}}Average\\ MCD\\ our\\ 1st\\ Approach\end{tabular}}} & \textbf{\begin{tabular}[c]{@{}l@{}}Average\\ MCD\\ our\\ 2nd\\ Approach\end{tabular}} \\ \hline
Set 1                                                                & 1.733                                                                        & \textbf{1.48}                                                                                              & 1.7                                                                                   \\ \hline
Set 2                                                                & 1.736                                                                        & \textbf{1.46}                                                                                              & \textbf{1.65}                                                                         \\ \hline
Set 3                                                                & 1.741                                                                        & 2.07                                                                                                       & 2.07                                                                                  \\ \hline
\end{tabular}
\caption{Speech synthesis test time results for subject 4 where audio is reconstructed from predicted MFCC 13, 100Hz}
\end{table}

\begin{table}[!ht]
\centering
\begin{tabular}{|l|l|l|}
\hline
\textbf{\begin{tabular}[c]{@{}l@{}}EEG\\ Feature\\ Set\end{tabular}} & \multicolumn{1}{c|}{\textbf{\begin{tabular}[c]{@{}c@{}}Average\\ MCD\\ our\\ 1st\\ Approach\end{tabular}}} & \textbf{\begin{tabular}[c]{@{}l@{}}Average\\ MCD\\ our\\ 2nd\\ Approach\end{tabular}} \\ \hline
Set 1                                                                & 1.211                                                                                                      & \textbf{1.14}                                                                         \\ \hline
\end{tabular}
\caption{Speech synthesis test time results for subject 1 where audio is reconstructed from predicted MFCC 128, 32Hz}
\end{table}

\begin{figure}[h]
\begin{center}
\includegraphics[height=8.5cm, width=\linewidth,trim={0.1cm 0.1cm 0.1cm 0.1cm}]{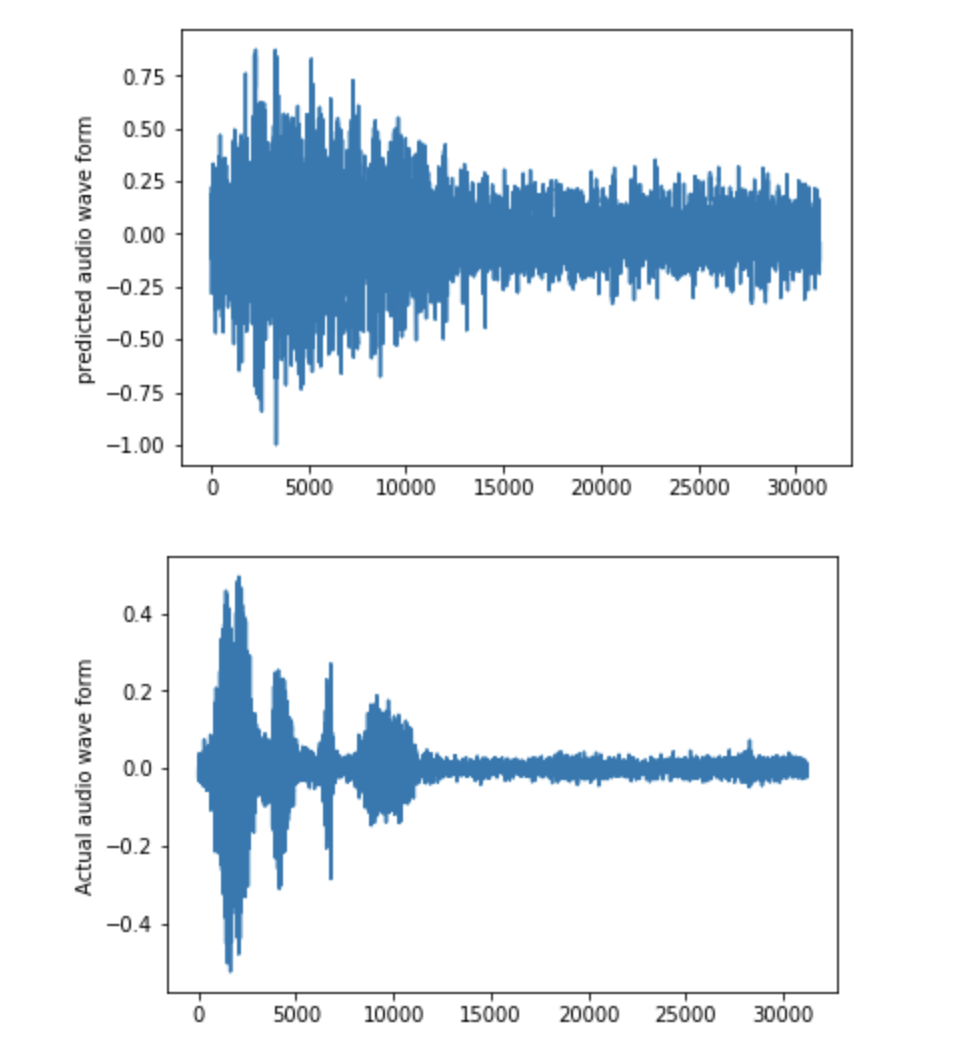}
\caption{Speech synthesis test time result for subject 1 where audio waveform is reconstructed from MFCC 128, 32 Hz with EEG Feature set 1 as input using \textbf{first} approach. The text corresponding to actual waveform was 'Hi Bixby'} 
\label{1vsall}
\end{center}
\end{figure}

\begin{figure}[h]
\begin{center}
\includegraphics[height=8.5cm, width=\linewidth,trim={0.1cm 0.1cm 0.1cm 0.1cm}]{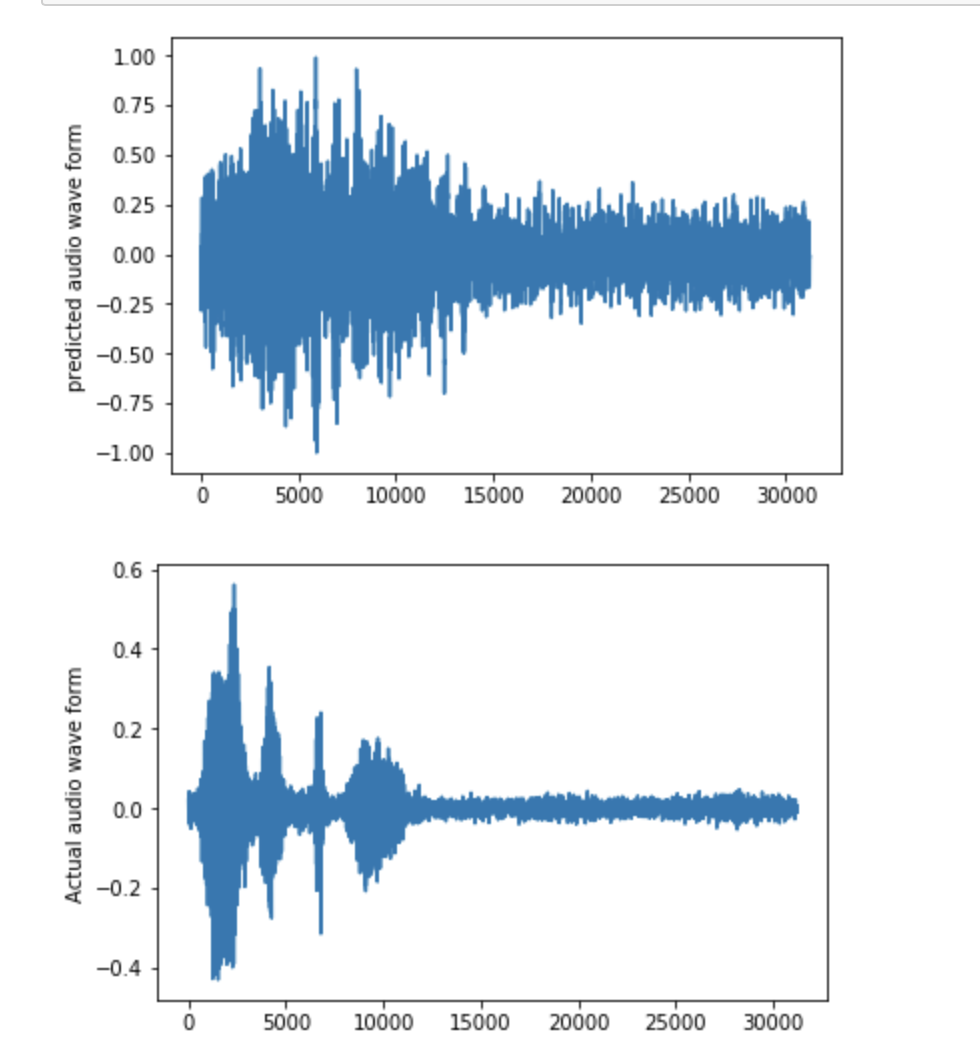}
\caption{Speech synthesis test time result for subject 1 where audio waveform is reconstructed from MFCC 128, 32 Hz with EEG Feature set 1 as input using \textbf{second} approach. The text corresponding to actual waveform was 'Hi Bixby'} 
\label{1vsall}
\end{center}
\end{figure}

\section{Conclusion and Future work}

In this paper we introduced attention-regression model to demonstrate predicting acoustic features from electroencephalography (EEG) features recorded in parallel with spoken sentences. First we demonstrated predicting acoustic features directly from EEG features and then we demonstrate predicting acoustic features from EEG features using a two-step approach where in the first step we use our model to predict articulatory features from EEG features and then in the second step another regression model is trained to transform the predicted articulatory features to acoustic features. 

Our proposed attention-regression model demonstrates superior performance compared to the regression model introduced by authors in \cite{krishna2020synthesis} when tested using their data set for majority of the subjects during test time. The results presented in this paper further advances the work described by authors in \cite{krishna2020synthesis}. We further conclude in this paper that using MFCC features of 128 coefficients sampled at 32 Hz as targets for the regression model results in generating more understandable audio from EEG features even though the test time mel cepstral distortion (MCD) value is slightly higher compared to using MFCC features of 13 coefficients sampled at 100 Hz as targets \cite{krishna2020synthesis}. 

There is a lot of scope for improving this work. Future work will focus on developing techniques to further improve the intelligible quality of the audio generated from EEG by generating audio waveforms from EEG with less noise. We believe our results can be improved by using EEG features recorded with higher signal-to-noise ratio.

\section{Acknowledgement} 
We would like to thank Kerry Loader and Rezwanul Kabir from Dell, Austin, TX for donating us the GPU to train the models used in this work.

\bibliographystyle{IEEEtran}

\bibliography{mybib}


\end{document}